\documentclass[12pt,preprint]{aastex}

\usepackage{graphics,graphicx}
\usepackage{amsmath}
\usepackage{natbib}
\usepackage{color}
\usepackage[percent]{overpic}
\def\apj{\rm ApJ}
\def\apjl{\rm ApJL}
\def\apjs{\rm ApJS}

\def\mnras{\rm MNRAS}
\def\nat{\rm Nature}
\def\pasj{\rm PASJ}

\def\aap{\rm AAP}

\newcommand{\gtorder}{\mathrel{\raise.3ex\hbox{$>$}\mkern-14mu
            \lower0.6ex\hbox{$\sim$}}}
\newcommand{\ltorder}{\mathrel{\raise.3ex\hbox{$<$}\mkern-14mu
            \lower0.6ex\hbox{$\sim$}}}

\shorttitle{SMBBH Population Estimates}
\shortauthors{}

\begin{document}

\title{Population Estimates for Electromagnetically-Distinguishable Supermassive Binary Black Holes}

\author{Julian H. Krolik\altaffilmark{1}}

\author{Marta Volonteri\altaffilmark{2}}

\author{Yohan Dubois\altaffilmark{3}}

\and

\author{Julien Devriendt\altaffilmark{4}}

\altaffiltext{1}{Department of Physics and Astronomy, Johns Hopkins University, Baltimore, MD 21218, USA}

\altaffiltext{2}{Institut d'Astrophysique de Paris, Sorbonne Universit\`e, CNRS UMR 7095,  98 bis bd Arago, F-75014 Paris, France}

\altaffiltext{3}{Institut d’Astrophysique de Paris, Sorbonne Universit\`e, CNRS, UMR 7095, 98 bis bd Arago, 75014 Paris, France}

\altaffiltext{4}{Department of Physics, University of Oxford, Keble Road, Oxford OX1 3RH, UK}

\begin{abstract}
Distinguishing the photon output of an accreting supermassive black hole binary system from that of a single supermassive black hole accreting at the same rate is intrinsically difficult because the majority of the light emerges from near the ISCOs of the black holes.  However, there are two possible signals that can distinctively mark binaries, both arising from the gap formed in circumbinary accretion flows inside approximately twice the binary separation.  One of these is a ``notch" cut into the thermal spectra of these systems in the IR/optical/UV, the other a periodically-varying excess hard X-ray luminosity whose period is of order the binary orbital period.  Using data from detailed galaxy evolution simulations, we estimate the distribution function in mass, mass ratio, and accretion rate for accreting supermassive black hole binaries as a function of redshift and then transform this distribution function into predicted source counts for these two potential signals.  At flux levels $\gtrsim 10^{-13}$~erg~cm$^{-2}$~s$^{-1}$, there may be $\sim O(10^2)$ such systems in the sky, mostly in the redshift range $0.5 \lesssim z \lesssim 1$.  Roughly $\sim 10\%$ should have periods short enough ($\lesssim 5$~yr) to detect the X-ray modulation; this is also the period range accessible to PTA observations.

\end{abstract}

\keywords{accretion, accretion disks; gravitational waves; galaxies: evolution; quasars: supermassive black holes}

\section{Introduction}\label{sec:intro}

Mergers of supermassive black hole pairs are widely expected to be both central to galaxy evolution and the dominant source of low-frequency gravitational wave emission in the Universe.  In the contemporary Universe, nearly every galaxy with luminosity comparable to or greater than the mean contains a supermassive black hole in its nucleus  \citep[][and references therein]{1998Natur.395A..14R,2013ARA&A..51..511K}, with many dwarf galaxies also hosting supermassive black holes \citep[][and references therein]{2013ApJ...775..116R,2017IJMPD..2630021M}.  Strong correlations between the masses of these black holes and the structures of their host galaxies \citep[e.g.,][]{2013ApJ...764..184M,2013ARA&A..51..511K} imply that the growth of these black holes is closely tied to the development of their hosts.  However, in cold dark matter cosmologies, successive mergers of smaller galaxies occur during the build-up of galaxies.
If galaxies acquire their supermassive black holes before the final round of mergers, there must be numerous opportunities for these black holes to form binaries, and, perhaps, ultimately merge.  When they do, the energy output is prodigious---several percent of $\sim 10^6 M_\odot$ amounts to $\sim 10^{59}$~erg---and it is emitted, in the form of gravitational waves, over timescales of only $\sim 10^3$~s.

Unfortunately, at the moment there are no confirmed examples of any supermassive black hole binaries; the best candidate so far is a radio galaxy with a pair of flat-spectrum cores $\approx 7$~pc apart \citep{Rodriguez2006}, whose relative motion, if interpreted as due to a binary orbit, suggests a mass $> 1.5 \times 10^{10} M_\odot$ \citep{Bansal2017}. Direct detection of their gravitational wave radiation awaits launch of a suitable space-based observatory, but the planned date for {\it LISA} remains fifteen years into the future.  There have been attempts to find such binaries through photon observations, but solid, credible examples have yet to emerge (for reasons we will discuss in a few paragraphs).

This situation is highly frustrating because finding examples of these systems at any stage, whether en route to merger as a binary, during the merger proper, or during the post-merger relaxation, would be of great interest.  The properties of supermassive black hole binaries as they evolve might shed considerable light on the way they interact with galaxy growth.   At the same time, establishing some idea of this population's distribution with respect to mass and mass ratio would provide significant aid to {\it LISA} mission planning. 

The fundamental difficulty in searching for them electromagnetically is that the light radiated by a binary black hole system should, in many respects, not differ greatly from that of a single black hole.  The majority of the power, both from the thermal disk and from the corona, emerges near the innermost stable orbit (ISCO), and if the scale of the ISCO is small compared to the binary separation $a$ (which is the case for the great majority of binaries because the timescale for orbital evolution by gravitational radiation losses is $\propto a^4$), this region is hardly influenced by the fact that the black hole has a binary partner.  Thus, to search for a binary means one must look for a characteristic involving only a small fraction of the emitted luminosity.

\subsection{Previous efforts to identify supermassive binary black holes using EM spectral signals}

Efforts to date have, for the most part, been focused on diagnostics that may apply to binaries, but with substantial uncertainties and caveats. One approach has been to search for periodicities in the continuum output, primarily on year to decade timescales \citep{Graham2015a,Graham2015b,LiuT2015,LiuT2016,Charisi2016,Dorn2017,Kova2018}.  Such a search requires considerable care because intrinsic AGN fluctuations generically have ``red" Fourier power spectra.  When the fluctuation power as a function of frequency $f$ is $\propto f^{-\alpha}$ and $\alpha > 1$ (as is usually the case for AGN variability), the integrated Fourier power, and therefore, the variance, diverge toward longer timescales.  As a result, quasi-periodic variations on timescales $\sim 1/3$ the duration of the monitoring almost always appear in lightcurves with this character \citep{Press1978}; because their timescale is a function of an experimental parameter, these quasi-periodicities are spurious, and disappear (to be replaced by longer-term apparent quasi-periodicities) when the data are extended (e.g., as reported by \citet{LiuT2018}).  Statistical analyses based on damped random walk (DRW) models (e.g., \citet{Graham2015b,LiuT2015,Charisi2016,Dorn2017}) fail to solve this problem because the DRW model assumes that the fluctuation power spectrum is flat at sufficiently low frequencies.  Although this fluctuation power flattening must eventually happen for any finite system, the timescale at which it does is only rarely reached in observed systems.  Searches for periodicity on these timescales (e.g., as proposed by \citealp{2007PASJ...59..427H,2009ApJ...700.1952H,Kelley2018}) also face the further obstacle that the prediction of periodicity is based upon either of two mechanisms, both of which face intrinsic difficulties.  One of these mechanisms rests upon the well-established periodic modulation of accretion from an outer circumbinary disk to the ``minidisks" around each of the members of a binary \citep{MM2008,Roedig2011,Shi2012,Noble2012}, but unless the binary separation is no more than a few times larger than the ISCO scale \citep{DAscoli2018}, the inflow time through the minidisks can be expected to be much longer than the period of the accretion modulation.  When this is the case, any such modulation is largely eliminated from the light emitted from the minidisk.  The other mechanism is the periodic Doppler boost due to the system's orbital motion.  For this mechanism to be detectable, the binary must be relatively compact and seen more-or-less edge-on (to produce a significant boost) and have a period short enough for $\gtrsim 5$--10 periods to be observed, as may be the case when the binary separation is quite small and the system is therefore rather short-lived \citep{Bowen2019}.

Another approach is to search for components within the broad emission line profiles that might indicate a binary.  Initial explorations of this method have in general been based upon a sum of two independent regions with a specific structure \citep{Bogdanovic2009,ShenLoeb2010,Nguyen2018}.  Unfortunately, as first pointed out by M. Penston and discussed briefly in \cite{Chen1989}, if gravity plays an important role in the emitting gas's dynamics, this method faces serious difficulties.  If the intrinsic spread in line-emitting gas velocity due to one black hole alone is $v_{\rm BLR}$, the line profiles of two accreting AGN blend when the orbital speed $v_{\rm orb} < v_{\rm BLR}$.  This is the situation when the binary separation $a \gtrsim r_i (1 + M_j/M_i)$, where $r_i$ is the size of the broad line region associated with black hole $i$, whose mass is $M_i$, and index $j$ refers to the other black hole.  In order for the displacement in velocity due to binary orbital motion to be sizable relative to the intrinsic spread in velocities, it is therefore necessary to reverse the inequality in distance scale.   However, when that is true, the binary separation must be comparable to or smaller than the intrinsic broad line region size unless there is a large mass contrast between the two black holes.  When that is the case, all the broad line gas is subject to the combined gravitational potential of both black holes and is also illuminated by the ionizing radiation from both black holes' accretion disks.  The result would be a combined broad line region, merged both with respect to dynamics and illumination.  If the binary separation is substantially smaller than the intrinsic scale of the line-emitting region, there would effectively be only a single broad line region, one whose potential is due to the sum of the two masses and whose ionizing luminosity is similarly the sum of the output of the two black holes.  If, on the other hand, the binary separation is comparable to the intrinsic line-emitting region scale, the combined potential would be qualitatively different from that of a single black hole AGN, and the way the gas is photoionized would also be substantially different from the single black hole situation \citep{ShenLoeb2010}.  In such a case, the emitted lines would not particularly resemble, whether in flux or velocity profile, any simple sum of two independent components.  Moreover, because so little is known with certainty about the source of gas for the broad line region or the mix of forces accounting for its internal motions, it is difficult to predict the character of a broad line region subject to the combined gravity and ionization radiation of two nearby AGN. 

Nonetheless, there have been many efforts to find SMBHBs by looking for line profiles that might be described as a sum of more conventional profiles.  Early work of this sort has been reviewed by \cite{Popovic2012}.  More recently, the focus has been on searches that also look for profile time-dependence possibly attributable to a binary in which only one of the two black holes is active \citep{Tsalmantza2011,Eracleous2012,Shen2013,JuGreene2013,LiuX2014,WangGreene2017,Runnoe2017}.  This approach avoids the problem of profiles superposed in velocity space, making it possible to search for smaller time-dependent offsets; however, it is difficult for this method to distinguish shifts due to orbital motion from shifts due to internal changes in the disposition of the emitting gas.  Similarly, it also avoids the problem of double illumination, but does not, however, mitigate the problem posed by a merged gravitational potential.  To date, several tens of candidates have been monitored for a number of years: insisting on rigorous follow-up before declaring them to be bona fide binaries, these searches have been able only to place upper bounds on the frequency of supermassive binary black holes (SMBBHs) detectable by this method.

\subsection{Proposed methods to identify supermassive binary black holes using EM spectral signals}

There are, however, other potential diagnostics of binarity that can be related in much more specific fashion to the binary nature of the system.   Two such were proposed by \cite{Roedig2014}.  Both arise from a distinctive dynamical property of accreting binaries in which $\sim 0.04 < q < 1$, where the mass ratio $q \equiv M_2/M_1$: a very low density gap is opened inside a radius from the binary center-of-mass $\approx 2a$ \citep{MM2008,Roedig2011,Shi2012,Noble2012,Dorazio2016}.  Narrow streams enter this gap from the inner edge of the circumbinary disk.  Matter with angular momentum close to the circular-orbit angular momentum at $r \approx 2a$ suffers strong positive torques from the binary and returns to the circumbinary disk; however, once it does, the shock that occurs upon striking the inner edge deflects some of the gas to lower angular momentum trajectories, and this gas falls inward \citep{Shi2015}, where it feeds a pair of ``minidisks", small accretion disks orbiting each of the binary's members.  The minidisks extend only out to their tidal truncation radii, $r_{T1,2} \approx 0.3a (q^{-0.3},q^{+0.3})$ \citep{Paczynski1977,Roedig2014}.  To make this system even more distinctively binary, the rate at which matter reaches the minidisks is strongly modulated on periods comparable to the binary orbital period \citep{MM2008,Shi2012,Farris2014,Shi2015}.

The first diagnostic created by this gap is a ``notch" in the thermal disk spectrum. Initial work on this topic \citep{Roedig2012,Tanaka2012,Gultekin2012,Kocsis2012} downplayed light from the minidisks, either because they were omitted from consideration or because they were thought to be dim.   However, more recent simulational work \citep{Ryan2017,DAscoli2018,Tang2018,Bowen2019} has confirmed the view of \citet{Roedig2014} that they can radiate with substantial luminosity. In an ordinary accretion disk, the light in the total disk spectrum at frequency $\nu$ is emitted primarily by the band of radii in the disk at which the temperature $T \sim h\nu/k_B$.  Because there is little gas in the gap, it cannot provide an optically thick surface covering its area to radiate at the temperatures that might have been found there if the binary were actually a single black hole of mass $M_1 + M_2$.  Moreover, the 3D MHD simulations of \citet{Shi2016} showed that such gas as there is is only weakly heated because, unlike the gas traversing stable circular orbits in a conventional accretion disk, its flow is laminar, and internal turbulent dissipation is very weak\footnote{2D HD calculations assuming that a phenomenological ``$\alpha$" viscosity acts in the stream merely because it has internal shear reach a different conclusion, as in \citet{Farris2015}}.  The energy that would be dissipated into heat while gas moves inward from a depth in the potential $\approx G(M_1+M_2)/2a$ to $ \approx GM_{1,2}/(0.3aq^{\pm 0.3})$ is instead converted into heat once the gas arrives at the outer edges of the minidisks. However, once gas joins a minidisk, its inward drift yields energy that is radiated thermally in the usual fashion. There is, in addition, some heat dissipated in the shock between torqued stream gas and the inner edge of the circumbinary disk. Thus, the overall result is little thermal output at the frequencies that would otherwise have been radiated from the radii in the gap, but (not quite) normal accretion disk emission at frequencies both lower and higher.  The characteristic energy of the missing photons is $k_B T_0$ for
\begin{equation}
T_0 \equiv \left(\frac{3}{8\pi \sigma} \frac{GM\dot M}{r^3}\right)^{1/4}/k_B \simeq 3.3 \times 10^4 \left[\dot m(\eta_{\rm acc}/0.1)^{-1} M_8^{-1} (a/100r_g)^{-3}\right]^{1/4}\hbox{~K}.
\end{equation}
Here $M$ and $\dot M$ are the total mass and accretion rate of the binary, $k_B$ is the Boltzmann constant, $\eta_{\rm acc}$ is the radiative efficiency, and $M$ is scaled to $10^8 M_\odot$.

The second diagnostic is the direct result of the shocks created when the accretion streams strike the outer edges of the minidisks.   When the binary separation is $\lesssim 3 \times 10^3 r_g$ (the binary gravitational radius $r_g \equiv GM/c^2$), the immediate temperature achieved in these shocks is $\gtrsim 100$~keV.  Electrons this hot readily cool by Compton upscattering the thermal seed photons created at inner radii of the minidisks to hard X-ray energies.  For two reasons, the X-rays radiated from these shocks may have a spectrum even ``harder" than those X-rays created by accretion disk coronae near the ISCO.  First, the ion temperature reached in these shocks is very high; second, the local mean intensity of seed photons is smaller than in the coronae by roughly the ratio $0.3a/r_{\rm ISCO})$.  In addition, the strong modulation of the accretion rate across the gap should be directly translated into an equally strong modulation of these hard X-rays because the Compton cooling time is quite short compared to the orbital timescale\footnote{As noted by the referee, the absence of the notch, or of any spectral hardening in the 0.3–7 keV X-ray band, has been used \citet{Foord2017} as a way to strengthen the ruling out of a binary candidate \cite{LiuT2018}}.

The goal of this paper is to quantify the number of such systems that might be observable.  We will do so by making use of the results of a galaxy evolution simulation with a large enough comoving volume, $(140\hbox{~Mpc})^3$ that, over its entire history, formed $\sim 10^4$ supermassive black hole binaries. The total number of black holes (BHs) that existed at any point in the simulation, i.e., that are given individual IDs, is a little over $4\times 10^5$. Some of them are ``lost'' to mergers, while others are ``lost'' because
their host galaxy has been disrupted.  The result is to bring the final number of BHs in galaxies at $z=0$ to $3.5\times 10^4$. This number includes all BHs in the same galaxy if multiple BHs coexist at a given time. This simulation also estimated the accretion rate onto these black holes, although its comparatively coarse spatial resolution ($\sim 1$~kpc) makes these estimates rather uncertain.  Despite this handicap, it does a reasonably good job of matching the observed luminosity function of luminous quasars, even while somewhat overestimating the quasar population at lower luminosities \citep{V2016}.  This portion of our approach resembles that of \citet{Kelley2018}.

From the data of this simulation, we first construct the distribution function with respect to total mass $M$, mass ratio $q \equiv M_2/M_1$, and accretion rate in Eddington units $\dot m$ for the rate at which supermassive black hole binaries are formed within a series of redshift slices.  The Eddington accretion rate is defined assuming a radiative efficiency of 10\%. Because the luminosity of both binary ``signatures" may be written as a function of these three parameters and the binary separation $a$, we can link the distribution function parameters to observed flux. The population follows from the distribution function for the creation rate multiplied by the system lifetime, but because the relevant luminosities are $\propto a^{-1}$, we can restrict our census to binaries with small enough separation that their evolution is driven by gravitational wave emission.  Finally, by computing an appropriate integral over the distribution function, it is possible to compute how many sources there are as a function of observed flux.  We complete the estimate by calibrating out the known bias of the simulation at bright fluxes.

\section{Calculational Details}\label{sec:calc}

\subsection{Counting supermassive black hole binaries in the galaxy evolution simulation}

To compute our predicted source count distribution for these two types of signals, we begin with the Horizon-AGN simulation \citep{2014MNRAS.444.1453D}.   The Horizon-AGN simulation is run with the Adaptive Mesh Refinement code {\sc ramses}~\citep{teyssier02} in a $\Lambda$CDM cosmology with total matter density $\Omega_{\rm m}=0.272$, dark energy density $\Omega_\Lambda=0.728$, amplitude of the matter power spectrum $\sigma_8=0.81$, baryon density $\Omega_{\rm b}=0.045$, Hubble constant $H_0=70.4$~km~s$^{-1}$~Mpc$^{-1}$, initial fluctuation power spectrum index $n_s=0.967$, all compatible with WMAP-7 cosmology~\citep{komatsuetal11}. The size of the box $L_{\rm box}=100 \, h^{-1} {\rm Mpc}$ with $1024^3$ DM particles, which gives a dark matter mass resolution of $8\times 10^7 \, \rm M_\odot$. From the level 10 coarse grid, a cell is refined (or unrefined) up to an effective resolution of $\Delta x=1$ proper-kpc (level 17 at $z=0$) when the mass in a cell is more (or less) than 8 times that of the initial mass resolution. The simulation includes prescriptions, described in more detail in~\cite{2014MNRAS.444.1453D}, for background UV heating, gas cooling including the contribution from metals released by stellar feedback, star formation and feedback from stellar winds, and type Ia and type II supernovae. 

Black holes with an initial seed mass of $10^5\, \rm M_\odot$ are created in cells where the gas and stellar density exceeds the threshold for star formation, $n_0=0.1\, \rm H\, cm^{-3}$, and where the stellar velocity dispersion is larger than $100\, \rm km\,s^{-1}$. To avoid multiple BHs forming in the same galaxy, an exclusion radius of 50 comoving-kpc is used. A dynamical drag mimicking dynamical friction exerted by gas on the black hole is included.  To estimate the accretion rate onto black hole, the simulation adopts a boosted Bondi-Hoyle-Lyttleton formalism \citep{Booth2009}, but the rate is capped at a rate that would produce an Eddington luminosity at a radiative efficiency of 10\%.  AGN feedback is a combination of two different modes, the so-called \emph{radio} mode operating when the luminosity falls below 1\% of the Eddington luminosity and the \emph{quasar} mode otherwise. In the quasar mode, 15\% of the AGN luminosity is injected isotropically into the heat content of the 4 cells surrounding the BH, while in the radio mode AGN energy is given to neighboring gas by a bipolar outflow with velocity $10^4\,\rm km\, s^{-1}$,  modeled as a cylinder of radius $\Delta x$ and height $2\Delta x$ weighted by a kernel function
\begin{equation}
  \label{eq:jetprofile}
  \psi\left(r_{\rm cyl}\right) = \frac{1}{2\pi\Delta x^2}\exp\left(-\frac{r_{\rm cyl}^2}{\Delta x^2}\right),
\end{equation}
with $r_{\rm cyl}$ the cylindrical radius. Mass is removed from the central cell and deposited in the cells enclosed by the jet at a rate $\dot{M}_{\rm J}$  proportional to the black hole mass accretion rate $\dot{M}_\bullet$:
\begin{equation}
  \label{eq:jetmass}
  \dot{M}_{\rm J}(r_{\rm cyl}) = \frac{\psi(r_{\rm cyl})}{\Psi} \eta_{\rm J} \dot{M}_\bullet
\end{equation}
where $\Psi$ is the integral of $\psi$ over the cylinder and $\eta_{\rm J} = 100$ is the mass-loading factor of the jet accounting for the mass entrained on unresolved scales (more details about the implementation are given in~\citealp{duboisetal10} and ~\citealp{2012MNRAS.420.2662D}). 

From this simulation we extract the merging black holes and create a population of binaries. In the simulation, black holes are merged when their separation falls below 4 cells ($4\times$1~kpc); to obtain more realistic properties for the binaries, we must therefore make some adjustments.  First, we assume that the actual mechanism of binary formation is for both black holes to sink to the galactic center as dynamical friction against the stars removes their orbital energy.   If they follow circular orbits and the stars form an isothermal sphere, this process takes a time \citep{1987gady.book.....B}: 
\begin{equation}
t_{\rm df}=0.672\, {\rm Gyr} \left(\frac{a}{4\, {\rm kpc}}\right)^2\left(\frac{\sigma}{100 \, {\rm km\, s^{-1}}}\right)\left(\frac{10^8 \,M_{\odot}}{M_2}\right)\frac{1}{\log(1+M_{\rm gal}/M_2)},
\end{equation}
where $\sigma$ and $M_{gal}$ are the central velocity dispersion and mass of the galaxy hosting the most massive black hole. $M_2$ is the mass of the lighter black hole, and we have multiplied the dynamical friction evolution timescale for circular orbits by a factor 0.3 to account for typical cosmological orbits being non-circular \citep{2003MNRAS.341..434T,2003ApJ...582..559V}. We have not included a correction to the initial $t_{\rm df}$ due to the change in mass of the host galaxy and the secondary black hole during that time.  We further suppose that dynamical friction continues to drive the evolution of the binary until additional processes, such as interaction with a circumbinary disk, become competitive. We do not include these additional processes here, but we note that they might introduce further delay \citep[see][for a review and additional references]{2014SSRv..183..189C}. For each binary we record the masses of the two black holes, $M_1$ and $M_2$, at the time when the merger ``starts", $t_{\rm in}$, i.e., when their separation falls below 4 cells and the simulation merges the black holes, as well as at the later time $t_{\rm in}+t_{\rm df}$.  The latter time defines the redshift $z$ at which we consider the binary to be formed.

\begin{figure}
\includegraphics[width=0.8\textwidth]{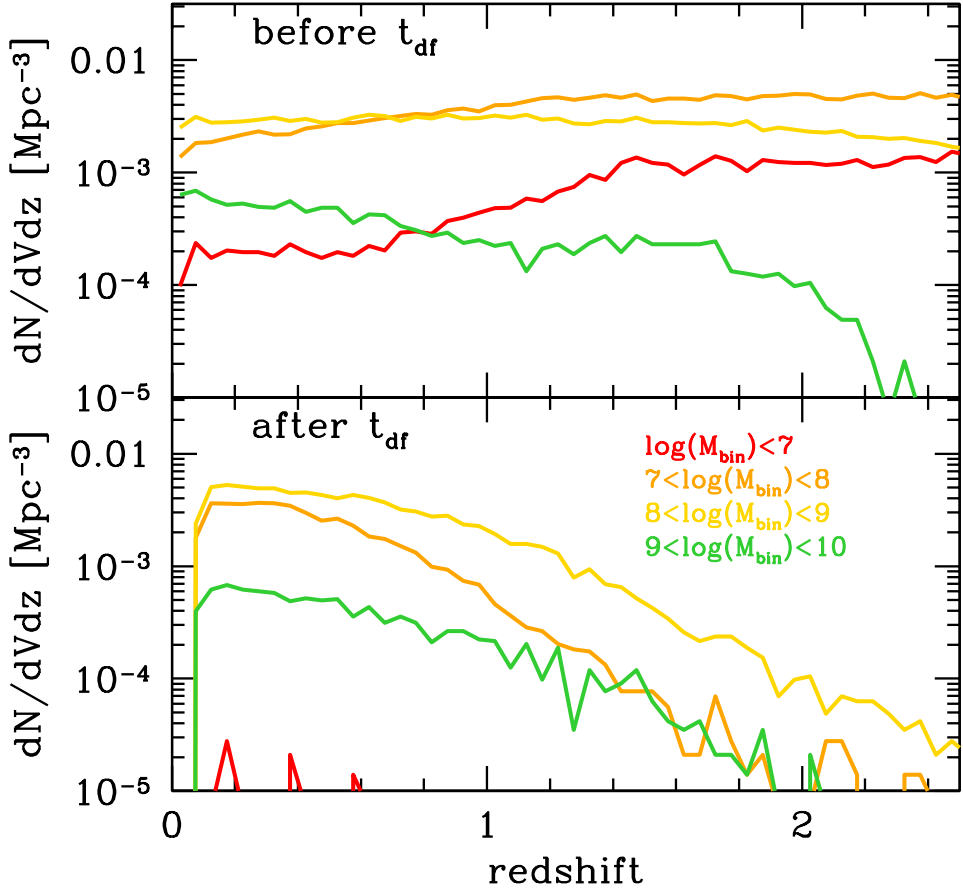}
\caption{Rate of binary production per unit comoving volume and redshift for four mass bins.  (Upper panel) Before allowance for the dynamical friction time-delay.  (Lower panel) After introduction of the dynamical friction time-delay.}
\label{fig:binary_production}
\end{figure}

We emphasize that the dynamical friction time-delay very substantially modifies the redshift distribution of the population we study (see Fig.~\ref{fig:binary_production}), with the effect increasing for smaller mass black holes.  For black hole masses $> 10^9 M_\odot$, the ``raw" creation rate per unit redshift and the ``delayed" rate are similar, both increasing steadily from $z\sim 2$ to the present, but the ``delayed" rate is diminished by a factor $\sim 2$ for $z \gtrsim 1.5$.  However, for black hole masses in the range $10^8$--$10^9 M_\odot$, although the ``raw" rate is roughly constant from $z=2.5$ to the present, after application of $t_{\rm df}$ the rate is reduced by an order of magnitude or more for all $z \gtrsim 1$, while the rate at $z \lesssim 0.5$ is augmented by a factor $\approx 2$--3.  In the range between $10^7$ and $10^8 M_\odot$, the ``raw" rate is also roughly constant as a function of $z$, but the ``delayed" rate is suppressed by more than an order of magnitude for $z > 1$.  Below $10^7 M_\odot$, the rate goes from comparable to the rate for the heaviest black holes to essentially zero.

Because $t_{\rm df}$ can be long, mass-growth by gas accretion over this time can be non-negligible, and the accretion rate will likely also differ significantly from that at  $t_{\rm in}$. We therefore use the simulation value for the total binary mass, $M$, at $t_{\rm in}+t_{\rm df}$, while we consider the mass ratio, $q= M_2/M_1$, unchanged from the beginning of the merger.  We tested that the results are not very sensitive to the mass ratio by using an alternative definition of $q = M_2/(M - M_2)$, where $M$ is the mass extracted from the simulation at $t_{\rm in}+t_{\rm df}$, while $M_2$ is the mass of the secondary at the earlier time $t_{\rm in}$. We also consider the accretion rate of the binary to be the rate onto $M$, and draw it directly from the simulation at $t_{\rm in}+t_{\rm df}$. Given the resolution of the simulation, it is reasonable to assume that this is an estimate of the mass available for accretion onto the binary, which is unresolved at 1~kpc resolution.

We calculated the gravitational wave strain at a frequency of 1~yr$^{-1}$ to compare with current upper limits on the stochastic background from Pulsar Timing Array  (PTA) experiments \citep{Shannon1522,doi:10.1093/mnras/stv1538,0067-0049-235-2-37}. We assumed circular binaries, and follow \cite{2001astro.ph..8028P} and \cite{2008MNRAS.390..192S} in calculating the background from a population of sources, to obtain a strain of  $1.89\times10^{-15}$, which is higher than the 95\% level upper limit of $10^{-15}$. As noted in the previous section and at the end of this section, the simulation is known to overestimate the quasar population. It should therefore also overestimate the binary population. We discuss at the end of this section how we deal with this bias.

The list of binaries is restricted to systems that form a binary by $z=0$, and all but 16 of these have final mass  $\geq 10^7 M_\odot$.  This limitation on the mass range is caused by our adding the dynamical friction timescale to the nominal creation time and insisting that $t_{\rm in}+t_{\rm df}$ is less than the age of the Universe at $z=0$: for low-mass black holes, the dynamical friction timescale is long; consequently, during the interval $t_{\rm df}$ after $t_{\rm in}$ they either grow to larger mass or reach a time past the age of the Universe.



From this list of binaries, we estimate the distribution function of black hole binary creation rate per unit volume with respect to mass, mass ratio, and accretion rate, $\partial^3 {\dot N}/\partial M\partial q \partial {\dot m}$, doing so within a set of redshift slices.  Because we wish to use this distribution function within an integral, we evaluate it on a grid.  In order to encompass the dynamic range of mass and accretion rate, it is evenly-spaced in $\log M/M_\odot$ and $\log {\dot m}$, while it is evenly-spaced in $q$.  The grid's range for $\log M/M_\odot$ is -6.5 to 9.5; for $\log {\dot m}$ it is -3. to 0.; for $q$ it runs from 0.01 to 1.0.  Labeling the cells in each parameters with $i$, $j$, and $k$, respectively,  we use a Gaussian kernel estimator \citep{Rosenblatt1956} to define the distribution function:
\begin{eqnarray}
&&{\partial^3 {\dot N} \over \partial \log M \, \partial \log {\dot m} \, \partial q} \left( \log M_i, \log {\dot m}_j, q_k\right) = {1 \over \pi^{3/2} V h_M h_{\dot m} h_q  \Delta t} \times \nonumber \\
&& \qquad\sum_n \exp\left\{-\left[(\log M_i - \log M_n)/h_M\right]^2 -\left[(\log {\dot m}_j-\log{\dot m}_n)/h_{\dot m}\right]^2 - \left[(q_k - q_n)/h_q\right]^2\right\}.
\end{eqnarray}
The sum is over all the entries (indexed with $n$) in the list of binaries created in the redshift slice.  For the results shown here, the smoothing lengths $h_M,h_{\dot m},h_q$ are all set to 0.25.  Their exact values make little difference provided the grid in each parameter is fine enough to resolve the associated smoothing length and the smoothing lengths chosen are not so great as to eliminate genuine structure in the simulation data (\citet{Rosenblatt1956} recommends smoothing lengths of order the expected gradient scales). This smoothing serves multiple purposes: it represents the uncertainties associated with the simulation data; it ensures that the distribution function varies on physically plausible scales rather than the much shorter scales that its sparse sampling would create without smoothing; and it reduces fluctuations due to small numbers of binaries in the less-populated regions of parameter space. The comoving volume of the simulation volume is $V$, while the duration of the redshift slice is written as $\Delta t$.

\subsection{Computing the source count distribution}

For thinking about the potential observability of these systems, the most useful population quantity is the source count distribution $dN/dF$.   In this context, the definition of ``observed flux $F$" needs some refinement because the features making these binaries distinctive pertain to {\it portions} of the luminosity, not the total luminosity.   Therefore, we define $F$ as the ``relevant" flux.  For the hard X-ray signal, it is the observed flux from the stream--minidisk shocks, whose luminosity we estimate by $({\dot m} L_E/\eta_{\rm acc}) (M)(r_g/a)$ because that is approximately the rate at which matter delivers kinetic energy to the shocks.  For the ``notch", the feature is {\it absent} flux, so its detectability must be measured in terms of the flux of the adjacent regions of the spectrum.  Moreover, the notch may be broad enough in wavelength that any single observation may detect only the long-wavelength cut-off or the short-wavelength recovery.  In order to discuss both cases in a unified manner, we---very crudely---estimate the relevant luminosity as the same luminosity available to the hard X-rays because that is the scale of the power {\it not} radiated thermally in the gap.

With this definition, the source count distribution for either signal due to a specific redshift slice is given by
\begin{eqnarray}\label{eqn:source_counts}
{dN \over dF}(z) = 4\pi r_{co}^2(z) \Delta r_{co} l_0^3& \int& \, d\log M \, \int \, d\log {\dot m} \, \int  \, dq \, \int_{a_{\rm min}/r_g}^{a_{\rm max}/r_g} \, d(a/r_g) \, \frac{dt}{da/r_g} \nonumber \\
&&\times {\partial^3 {\dot N} \over \partial \log M \partial \log {\dot m} \partial q} \delta\left[F - \frac{\dot m L_E(M)}{4\pi D_L^2 \eta_{\rm acc}}\frac{r_g}{a}\right],
\end{eqnarray}
where
\begin{equation}
{dt \over da/r_g} = (5/64)(a/r_g)^3 [(1+q)^2/q] (r_g/c)
\end{equation}
when the binary's evolution is due entirely to gravitational radiation.  The volume of the slice is determined by its dimensionless co-moving radius $r_{co}$ and the Hubble length $l_0$.

The $\delta$-function can be used to evaluate the integral over $a/r_g$.   Its zero is found at
\begin{equation}
\frac{a_*}{r_g} = \frac{ \dot m L_E(M)}{4\pi D_L^2 \eta_{\rm acc}F},
\end{equation}
but integrating with respect to $a/r_g$ introduces a multiplicative factor $\propto (a/r_g)^2$.   If $a_*/r_g$ lies outside the permitted range, the local integrand becomes zero.   Because the time per change in separation is itself $\propto (a/r_g)^3$, the integral becomes
\begin{eqnarray}
{dN \over dF}(z) =4\pi  r_{co}^2(z) \Delta r_{co} l_0^3 &\int& \, d\log M \, \int \, d\log {\dot m} \, \int  \, dq \, 
{\partial^3 {\dot N} \over \partial \log M \partial \log {\dot m} \partial q} \nonumber \\
&\times& (5/64)(a_*/r_g)^5 [(1+q)^2/q] (r_g/c) (4\pi D_L^2 \eta_{\rm acc})/\left[\dot m L_E(M)\right].
\end{eqnarray}
Writing $L_E(M) = L_{E\odot} {\cal M}$ and $r_g = r_{g\odot}{\cal M}$ for ${\cal M}$ the binary mass in $M_\odot$ units, the result is
\begin{eqnarray}
{dN \over dF}(z) = \frac{5}{4}\pi^2 r_{co}^2(z) &\Delta r_{co} l_0^3& \frac{r_{g\odot}}{c} \frac{D_L^2 \eta_{\rm acc}}{L_{E\odot}} \int \, d\log M \, \int \, d\log {\dot m} \nonumber\\
&\times& \int  \, dq \, {\partial^3 {\dot N} \over \partial \log M \partial \log {\dot m} \partial q} (a_*/r_g)^5 \frac{(1+q)^2}{q\dot m}.
\end{eqnarray}
This is computationally the most efficient form to use because one needs to find $a_*/r_g$ explicitly in order to compare it to the limits on $a/r_g$, but there is no need to present its dependence on $M$ and $\dot m$ explicitly.  On the other hand, it can also be conceptually useful to do so.   In this form, we have
\begin{eqnarray}\label{eq:concept}
{dN \over dF}(z) = \frac{5}{4^6\pi^3} r_{co}^2(z) &\Delta r_{co} l_0^3& \frac{r_{g\odot}}{c} \left(\frac{L_{E\odot}}{D_L^2 \eta_{\rm acc} F} \right)^4 F^{-1} \int \, d\log M \, \int \, d\log {\dot m}\nonumber \\
&\times& \int  \, dq \, {\partial^3 {\dot N} \over \partial \log M \partial \log {\dot m} \partial q} \frac{(1+q)^2}{q} M^5 {\dot m}^4.
\end{eqnarray}
What is learned from this form is that there is a very steep overall dependence on flux, $\propto F^{-5}$, but it can potentially be curbed by the equally steep, but positive, dependence on $\dot m$ and $M$.

Once the source counts associated with an individual redshift slice are computed, the total $dN/dF$ can be found simply by summing them because each represents the number of objects in the entire sky within a range of redshifts.  What we will present in the next section is the cumulative distribution, $N(>F) \equiv \int_F \, dF^\prime dN/dF^\prime$.

The distinction between the X-ray counts and the ``notch'' counts enters in the determination of the limits on $a/r_g$.   We impose absolute limits in both cases.  On the one hand, $a_{\rm min}/r_g$ can be no smaller than 15 because thinking in terms of distinct minidisks is no longer appropriate when the binary separation is so small---the outer edges of the minidisks are not that far outside their ISCOs.   On the other, we accept no values of $a/r_g$ larger than 1000 because the available energy for these two binarity signals becomes very small and because at larger separations it is increasingly unlikely that gravitational radiation controls the lifetime of the system. In particular, at larger separations accretion may influence the orbital evolution of binaries. Unfortunately, its net effect remains unclear---even its sign is debated \citep{MM2008,Shi2012,Tang2017,Miranda2017,Munoz2019,Moody2019}---but its characteristic timescale can become comparable to or shorter than the gravitational radiation timescale when $M_7\dot m (a/1000r_g)^4 \gtrsim 10$ \citep{2009ApJ...700.1952H,2019MNRAS.482.4383F}.    Because we choose $a/r_g \leq 1000$ and the great majority of our objects have $\dot m \lesssim 0.1$, this condition is satisfied by an extremely small fraction of our sample. The upper limit on $a/r_g$ is associated with a lifetime at that separation $\simeq 1.3 \times 10^6 [(1+q)^2/q] M_8 (a/10^3r_g)^4$~yr; the lifetime at the lower limit is only $22[(1+q)^2/q] M_8 (a/15r_g)^4$~d.  These are the only limits on $a/r_g$ applied to the X-ray counts.

However, the notch situation is more complex.  Here, another condition applies in addition to the same absolute separation limits imposed in the X-ray case.  To see this signature of binarity, {\it either} the low-energy cut-off or the high-energy recovery must appear in the NIR/optical/UV band; \citet{Roedig2014} find that these features are found at rest-frame photon energies $\approx kT_0$ and $10kT_0$, respectively.  Consequently, we define the feature as ``observable" if {\it either} $kT_0/(1+z)$ or $10kT_0/(1+z)$ lies between 1 and 10~eV (i.e., 1.2~$\mu$m and 1200~\AA).  With this definition, it is essentially impossible for both to fall within this range.  Taking the scaling for disk surface temperature applicable when the boundary condition and relativistic correction factors are close to unity (i.e., $T \propto r^{-3/4}$), there is a contribution to the counts whenever the inequality
\begin{equation}
8.7 \times 10^3 [Q/(1+z)]^{4/3} (\dot m/M)^{1/3} < a/r_g < 1.9 \times 10^5 [Q/(1+z)]^{4/3} (\dot m/M)^{1/3} \end{equation}
is satisfied for {\it either} $Q =1$ or $Q=10$. 

The final step in our procedure is to adjust both source count distributions for the known biases in the Horizon-AGN simulation which, as noted above, overestimates the luminosity function of AGN in some luminosity/redshift ranges.  To do so, we compute the ratio of our predicted binary source counts to the AGN source counts also predicted by this simulation.  We then apply this ratio to the {\it observed} AGN source counts as measured in the {\it Chandra} deep field \citep{Lehmer2012}.  The result is our best estimate of the actual source counts for our two diagnostics.

\subsection{The orbital period and redshift distributions as functions of flux}

In addition to this prediction of how many of these systems might be seen at observable flux levels, it is also of interest to predict their distributions in observed orbital period and redshift: both are potentially measurable.  Our formalism for predicting both distributions is very similar conceptually to the one used for the source counts, but differs in detail.

We begin with the orbital period distribution, for which the quantity of interest is $\partial^2 N/\partial F\partial P_{\rm orb}$ rather than $dN/dF$.  The integral of Eqn.~\ref{eqn:source_counts} is therefore subject to an additional $\delta$-function, $\delta\left[P_{\rm orb} - 2\pi (1+z) (a/r_g)^{3/2} r_g/c\right]$.  If the $\delta$-function in flux is applied to the integral over $a/r_g$ as done before, the $\delta$-function in period must then be applied to the integral over $\log M$.  To evaluate both integrals over the $\delta$-functions, one uses the form
\begin{equation} 
\int \, ds \, \int \, dt \, \delta[x - f(s,t)] \delta[y - g(s,t)] = \int \, ds \, \int \, dt \, \delta[s - s_*(x,y)] \delta[t - t_*(x,y)] |\partial (x,y)/\partial(s,t)|^{-1},
\end{equation}
where $s_*$ and $t_*$ are the solution to the joint constraints imposed by the $\delta$-functions, and $|\partial (x,y)/\partial(s,t)|$ is the determinant of the Jacobian.   The result is then
\begin{equation}
{\partial^2 N \over \partial F\partial P_{\rm orb}}(z) = \frac{0.43 \pi}{4(1+z)} r_{co}^2(z) \Delta r_{co} l_0^3 \frac{r_{g\odot}}{c} \frac{D_L^2 \eta_{\rm acc}}{L_{E\odot}} \int \, d\log {\dot m} \int \, dq \, 
{\partial^3 {\dot N} \over \partial \log M \partial \log {\dot m} \partial q} (a_*/r_g)^{7/2} \frac{1}{r_{g*}} \frac{(1+q)^2}{q\dot m},
\end{equation}
where $r_{g*}$ is, like $a_*/r_g$, the value picked out by the $\delta$-function, and the distribution function is evaluated at $\log M_*$.  In this case, $a_*/r_g = \left(\dot m P_{\rm orb}/2\pi \eta \kappa_T F\right)^{2/5} (c^2/D_L)^{4/5}$ (subject to the cut-offs) and $GM_* = \left[ (P_{\rm orb}D_L)^2 (c \eta \kappa_T F)^3/(2\pi)^2\right]^{1/5}$.
The quantity $\partial^2 N/\partial F\partial P_{\rm orb}$ may then be interpreted as the probability density with respect to observed orbital period at fixed flux.

We define the redshift distribution for fixed flux in very similar fashion.  It is
\begin{equation}
\frac{df}{dz}(F) \equiv \left[\frac{\partial^2 N}{\partial \ln F \partial z}\right] / \left[\frac{dN}{d\ln F}\right].
\end{equation}
It is particularly easy to compute from our data because our lists of binaries are compiled for separate redshift slices.  The above ratio can then be computed by taking the ratio of the source count distribution in each slice to the total source count distribution.

\section{Results}

To estimate the binary creation rate distribution function, we divided the simulation's list of binaries into redshift slices with width $\Delta z = 0.2$, from $z=2.0$ to $z=0.4$.  From $z=0.4$ to $z=0$, the slices were finer, $\Delta z = 0.1$.  Doing so yielded several thousand binaries in each slice, enough to define a distribution function reasonably well, while keeping the slices narrow enough that the approximation of placing all binaries at the same luminosity distance (corresponding to the central redshift of the slice) is reasonable.

\begin{figure}
\centering
\includegraphics[width=0.44\textwidth,angle=90]{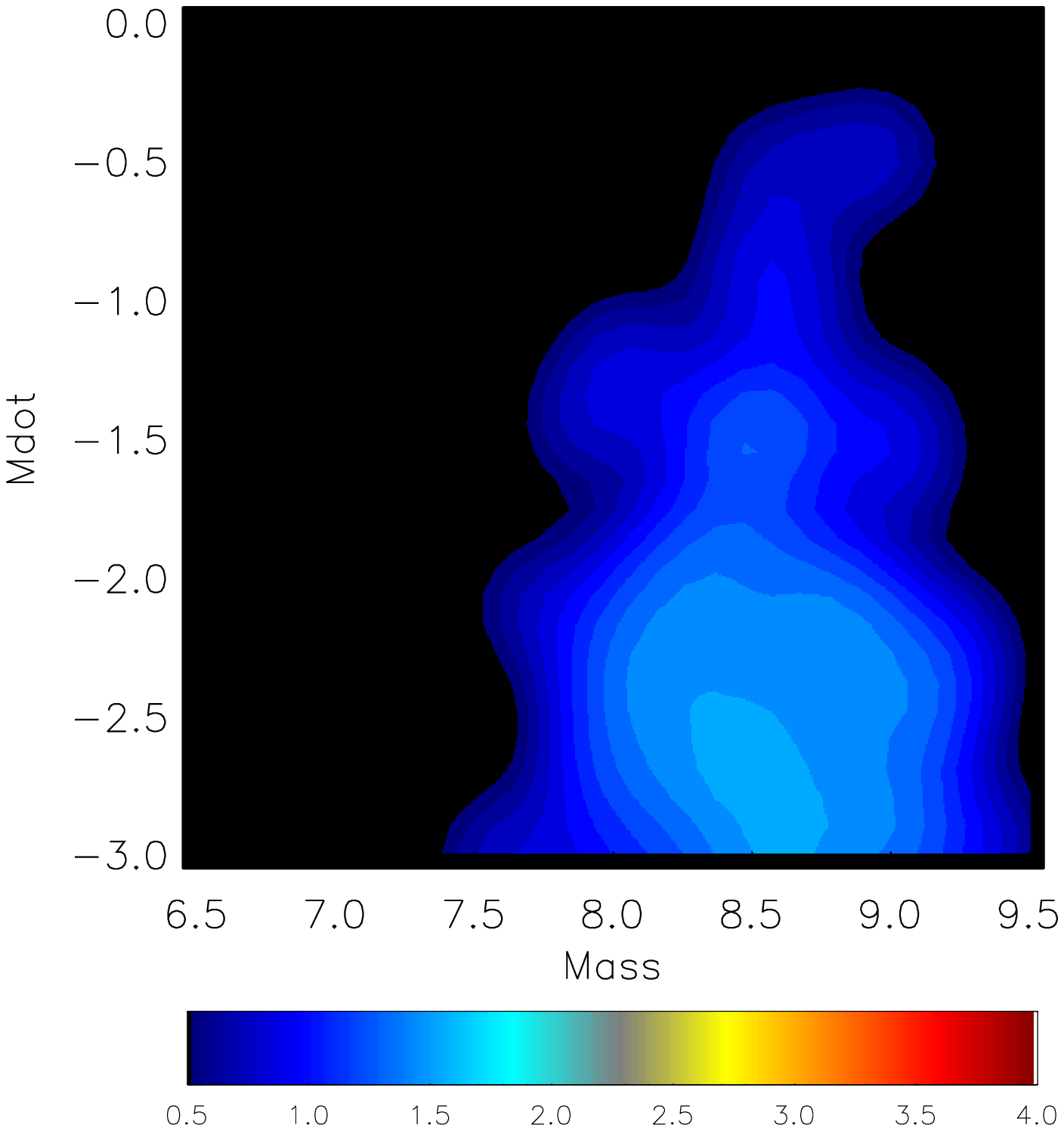}
\hskip -38mm
\includegraphics[width=0.44\textwidth,angle=90]{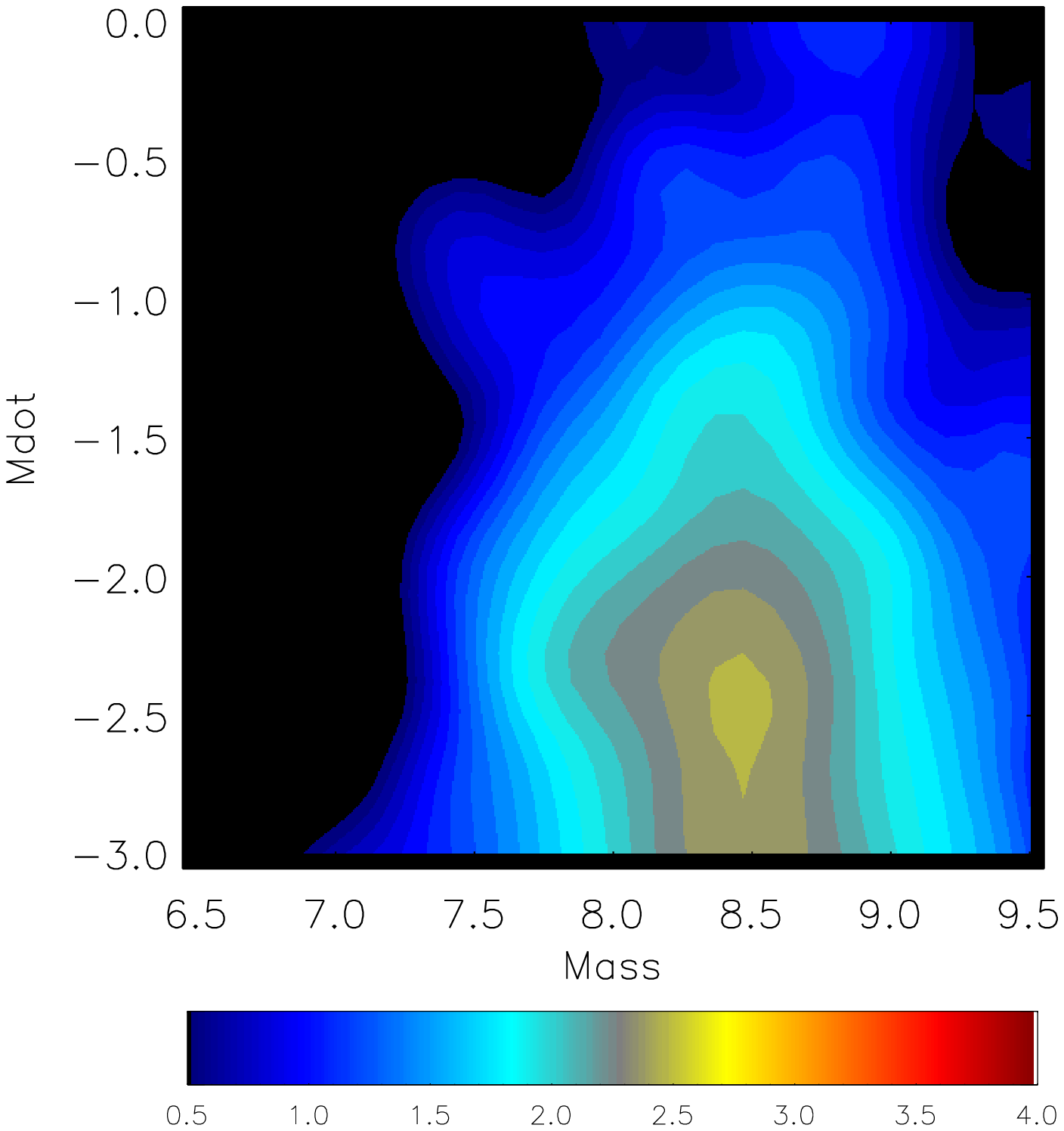}
\includegraphics[width=0.44\textwidth,angle=90]{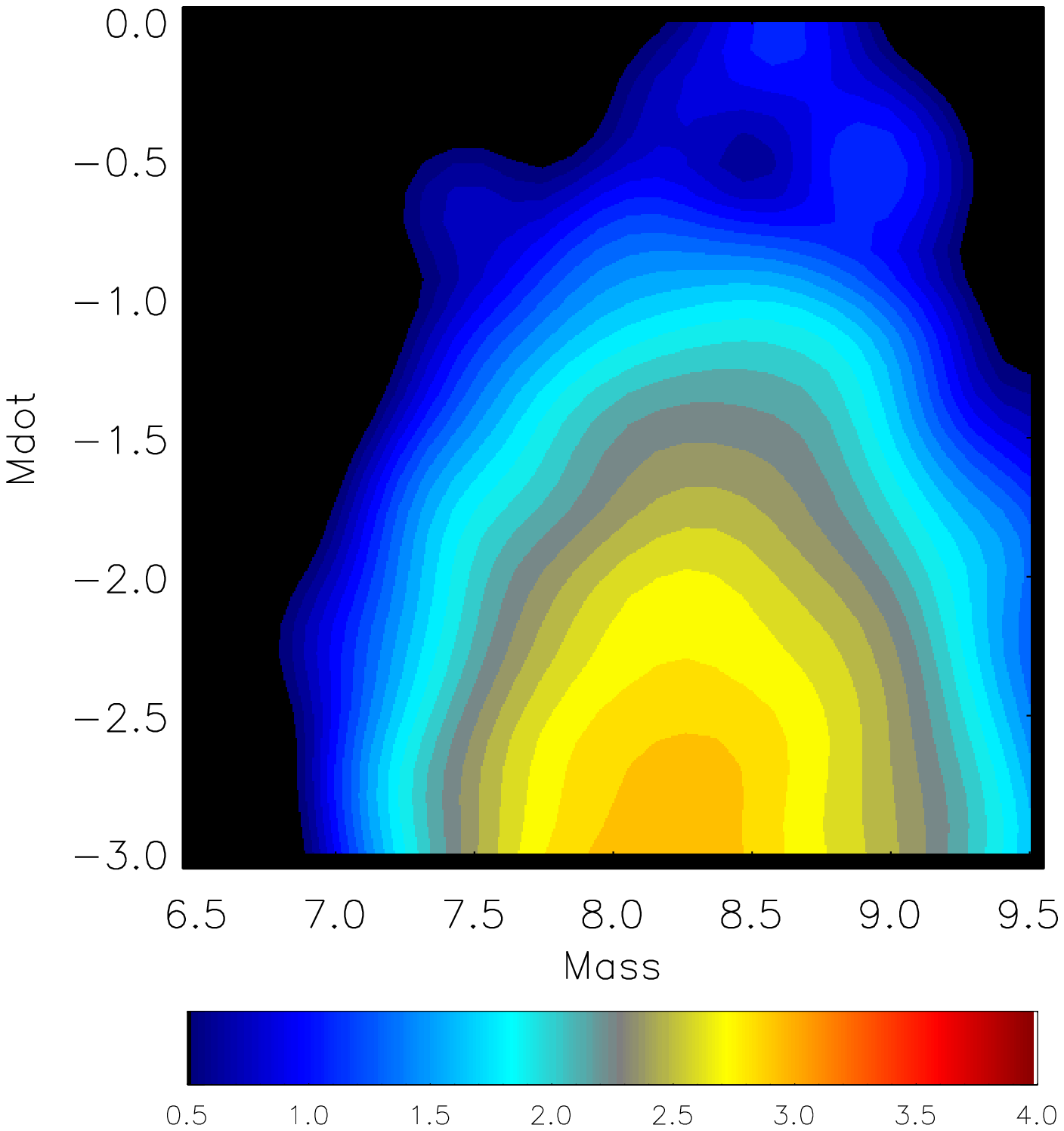}
\hskip -38mm
\includegraphics[width=0.44\textwidth,angle=90]{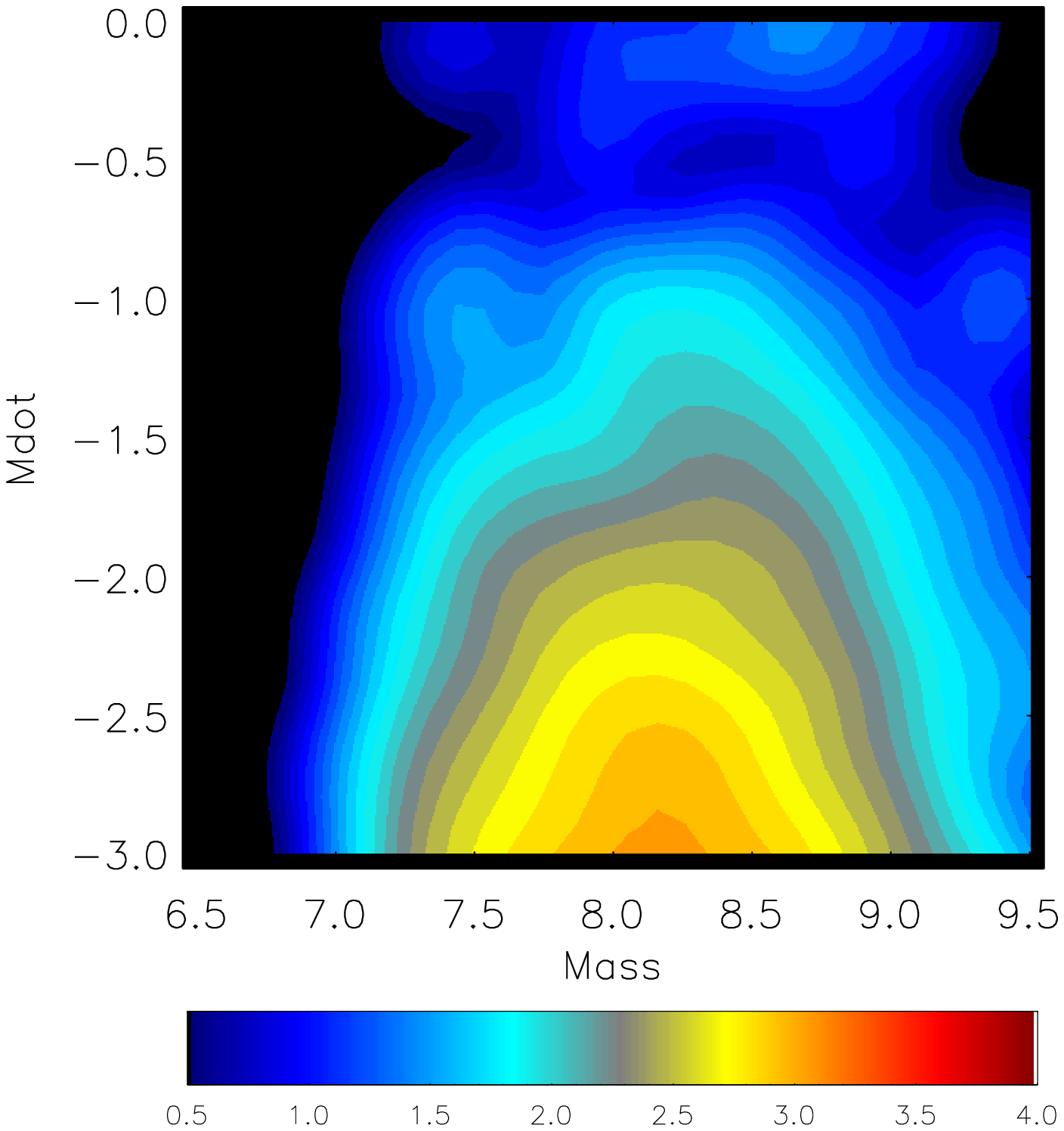}
\caption{Distribution function of binary creation rate per $10^6$~Mpc$^3$ (co-moving) per the time associated with $\Delta z$ for the redshift slice, per $\log M$ per $\log \dot m$ (i.e., integrated over mass-ratio).  The contours are logarithmic in number density, and the colorscale is uniform across all panels.  From left to right and then top to bottom, these show the redshift slices 1.8--2.0, 1.2--1.4, 0.6--0.8, and 0.2--0.3.}
\label{fig:distfctn}
\end{figure}

Because we have included a lengthy delay for the actual formation of the binary (the dynamical friction time $t_{\rm df}$), the total number of binaries per unit redshift increases steadily from $z=2$ until it reaches a roughly constant rate between $z=0.4$ and $z=0.1$, and drops thereafter (see also Fig.~\ref{fig:binary_production}).  The number per redshift at the time of maximum production is $\sim 3\times$ the number at $z=1$.  However, as shown in Figure~\ref{fig:distfctn}, the mean mass of the binaries falls slightly over time, declining from $\simeq 10^{8.6} M_\odot$ at $z=2$ to $\simeq 10^{8.2} M_\odot$ by $z=0.4$, while the breadth of the mass distribution widens.  At the same time, the accretion rate distribution shifts downward by roughly an order of magnitude.  Because luminosity is $\propto M \dot m$, the net result is to place more binaries at low redshift, while they are intrinsically brighter at higher redshift.

\begin{figure}
\begin{center}
\includegraphics[width=0.44\textwidth,angle=90]{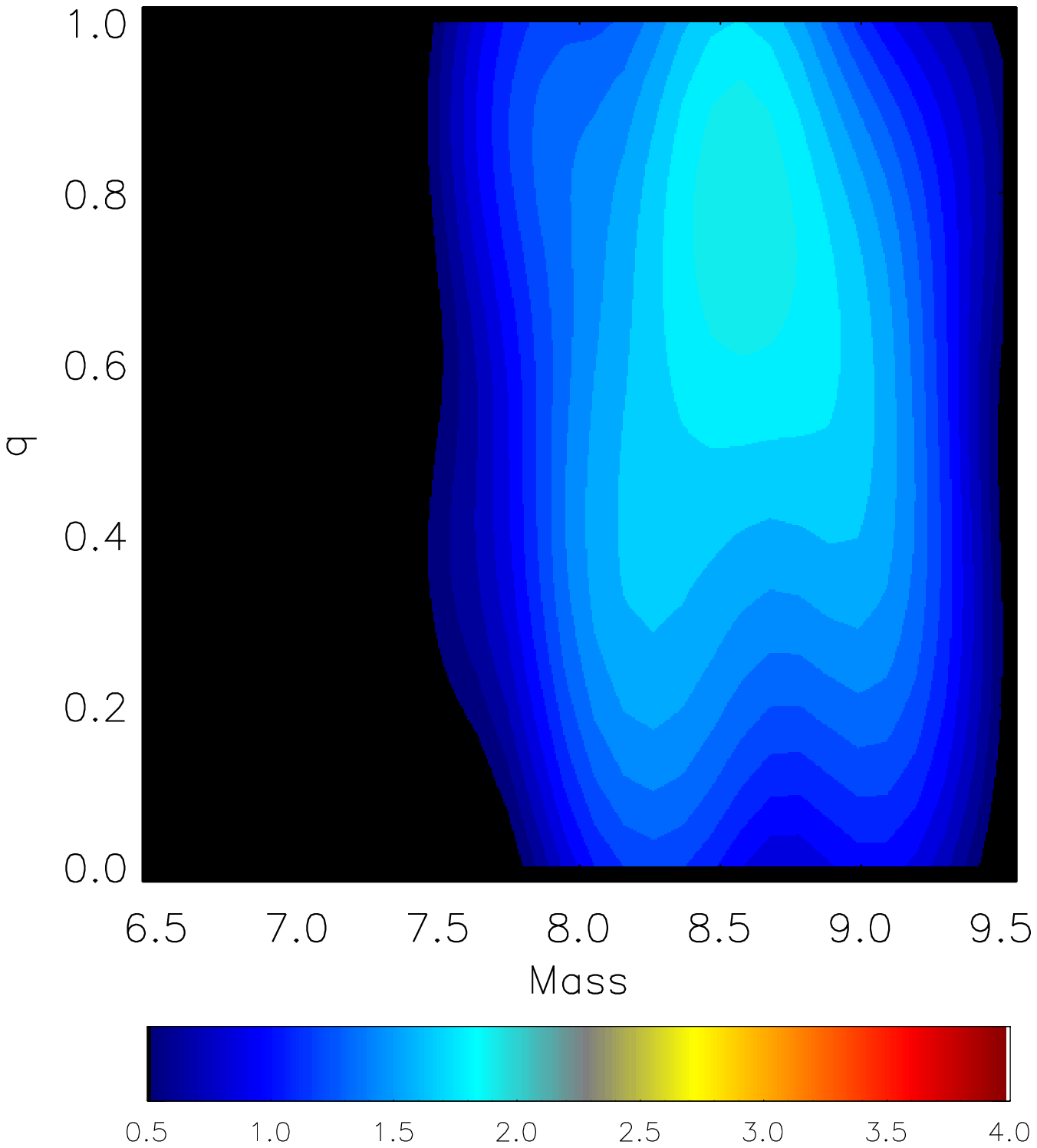}
\hskip -38mm
\includegraphics[width=0.44\textwidth,angle=90]{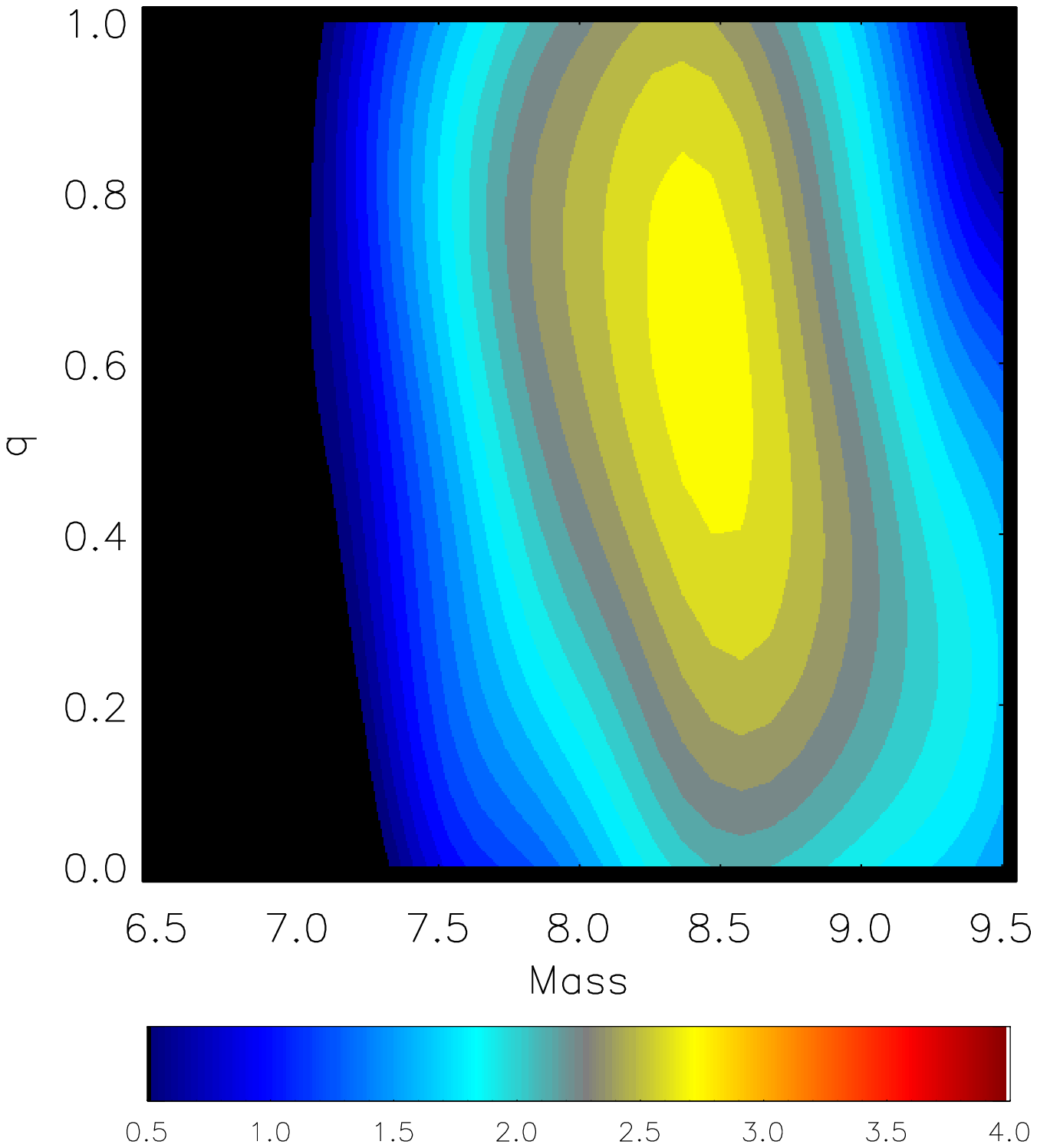}
\includegraphics[width=0.44\textwidth,angle=90]{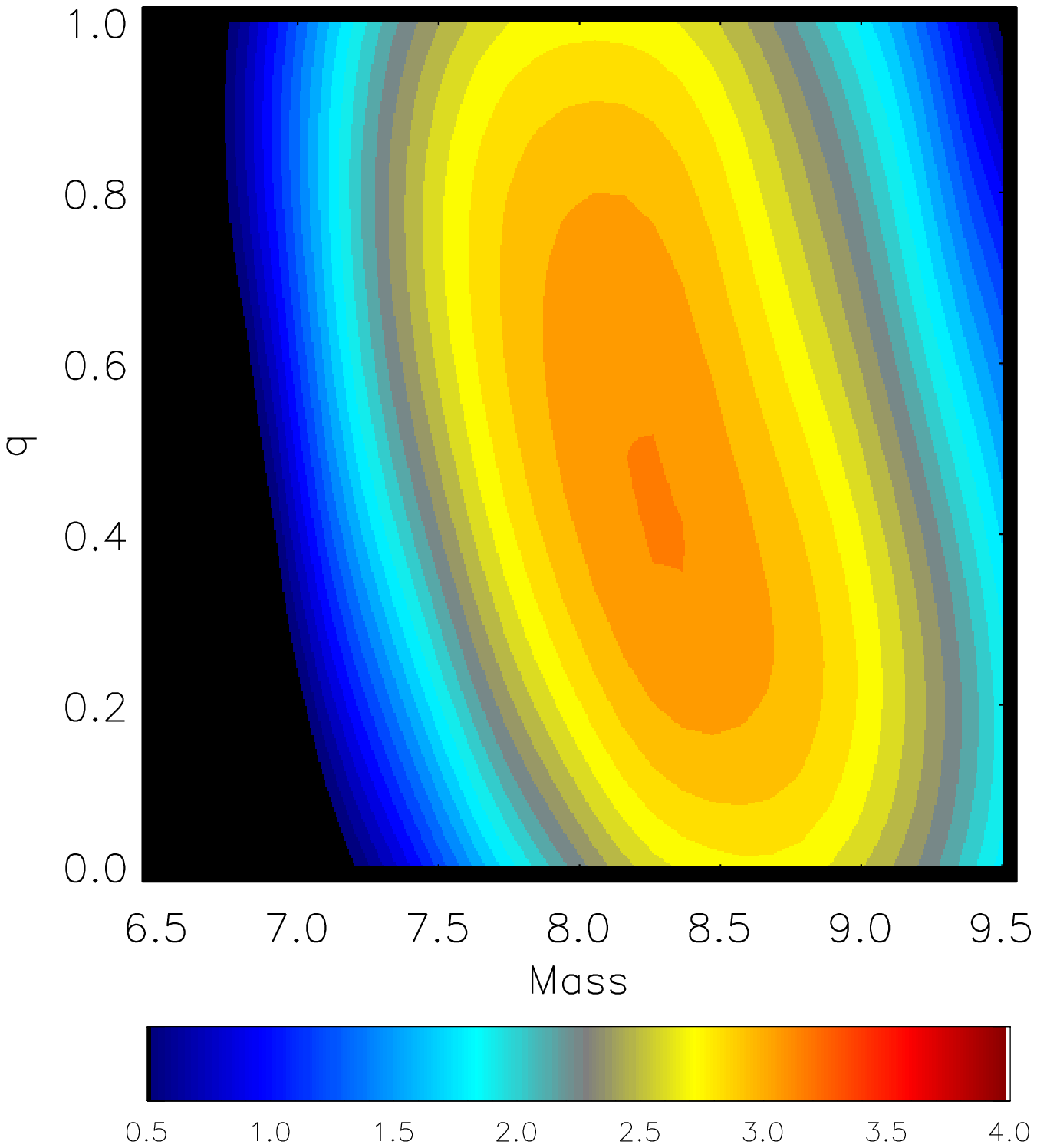}
\hskip -38mm
\includegraphics[width=0.44\textwidth,angle=90]{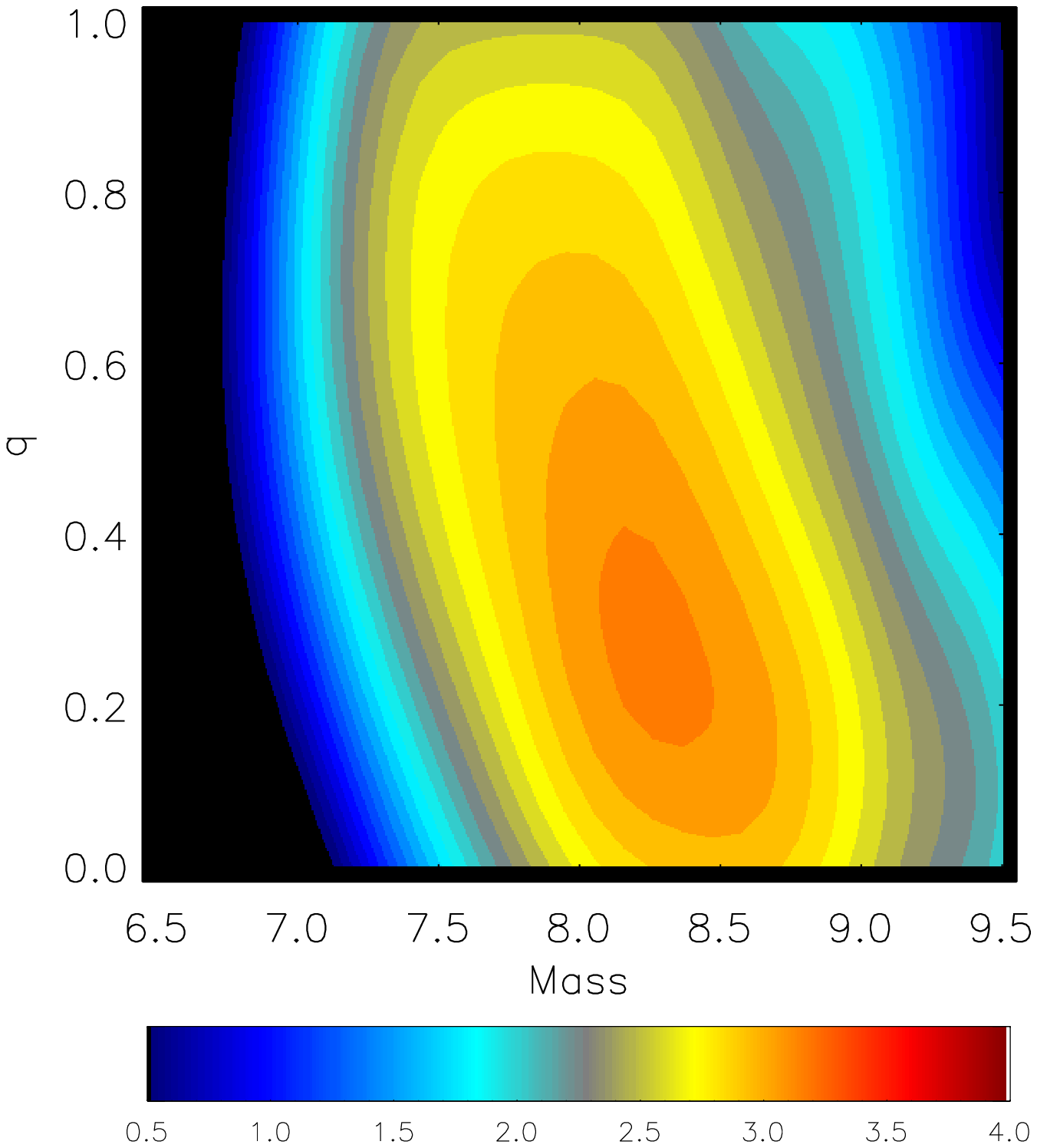}
\caption{Distribution function of binary creation rate per $10^6$~Mpc$^3$ (co-moving) per the time associated with $\Delta z$ for the redshift slice, per $\log M$ per $q$ (i.e., integrated over $\dot m$).  The contours are logarithmic in number density.  From left to right and then top to bottom, these show the redshift slices 1.8--2.0, 1.2--1.4, 0.6--0.8, and 0.2--0.3.}
\label{fig:qdistfctn}
\end{center}
\end{figure}

At the highest redshifts we simulated, mass-ratio and mass are almost independent, but for $z \lesssim 1$, there is a strong anticorrelation between mass and mass-ratio (Fig.~\ref{fig:qdistfctn}).  At these later times, binaries with lower masses have mass-ratios close to unity, while in higher-mass systems $q$ is more likely to be $\approx 0.1$--0.2.  The trend in between is almost linear in terms of $q$ vs. $\log M$ and steepens with decreasing redshift.  Because the total mass of the binary is dominated by the primary, this trend suggests that if a binary is formed with one black hole of mass $M$, the probability distribution for the other mass reflects the mass function for all supermassive black holes, in which smaller masses outnumber larger masses.

\begin{figure}
\includegraphics[width=0.6\textwidth,angle=90]{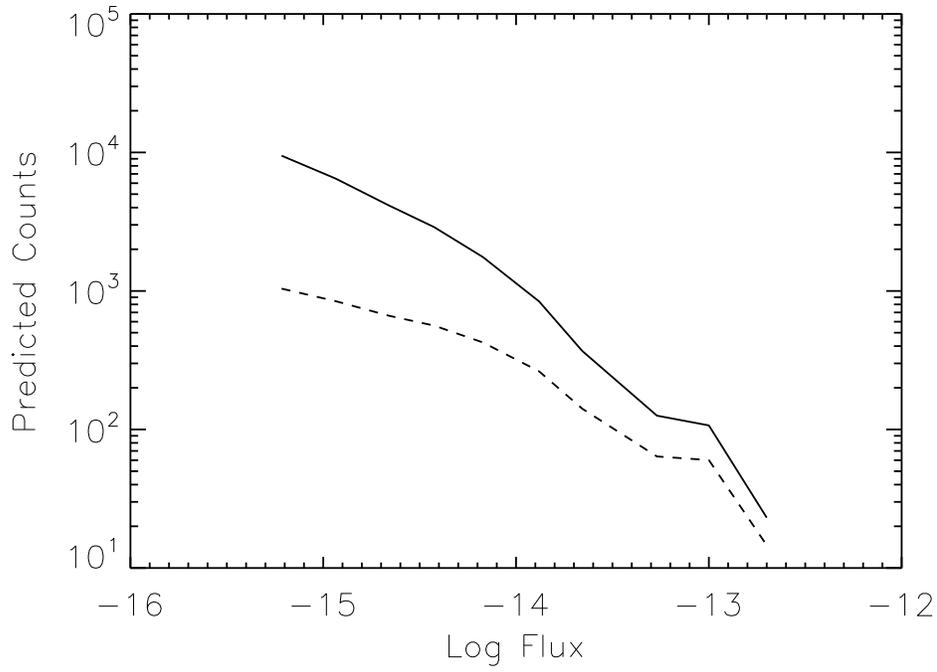}
\caption{Predicted source counts $N(>F)$ for the hard X-ray binary signature (solid curve) and the ``notch" signature (dashed curve).}
\label{fig:counts}
\end{figure}

These distribution functions lead to the cumulative source count functions $N(>F)$ shown in Figure~\ref{fig:counts}.  In evaluating these predicted source counts, it is important to recognize that they should be given sizable uncertainty.  Both the simulations and the dynamical arguments supporting the two binarity signatures contain significant systematic errors.  We do not show numbers for fluxes greater than $2 \times 10^{-13}$~erg~cm$^2$~s$^{-1}$ because statistical errors (in both the simulation data and the existing all-sky X-ray surveys) become too large.  We likewise cut off the predicted distribution at $5 \times 10^{-16}$~erg~cm~s$^{-2}$ because both the simulation data and the normalizing observational data have larger errors at lower fluxes.

Nonetheless, as these curves show, even at fluxes above $\sim 10^{-13}$~erg~cm$^{-2}$~s$^{-1}$ we expect significant numbers---$\sim 10^2$---of detectable supermassive binary black holes to exist, even though they represent only a small fraction of the total population of accreting black holes ($\sim 10^{-4}$ for X-ray flux $\sim 10^{-13}$~erg~cm$^{-2}$~s$^{-1}$).  The predicted numbers of systems with both signatures, hard X-ray and ``notch", are very similar for the higher flux sources because for those fluxes the wavelengths of the notch edges are nearly always in the observable range; consequently, the primary requirement for observability of the ``notch" and the hard X-ray component is the flux associated with either feature.  At lower fluxes, the predicted population of ``notches" falls below that of the hard X-ray cases, but by less than an order of magnitude even at $\sim 1 \times 10^{-15}$~erg~cm$^{-2}$~s$^{-1}$.

In the high-flux range where the two curves are close, $d\ln N/d\ln F$ falls from $\simeq -2$ for $F > 10^{-13}$~erg~cm$^{-2}$~s$^{-1}$ to $\simeq -0.9$ between $10^{-13}$ and $10^{-14}$~erg~cm$^{-2}$~s$^{-1}$.  The slope becomes even shallower at lower fluxes, particularly for the notch signal.  A progressively shallower slope for $N(>F)$ as $F$ decreases would result from cosmological kinematics alone if the luminosity function per comoving volume were constant: the co-moving radius per unit redshift declines with increasing redshift (it is $\propto (1+z)^{-3/2}$ in a flat and matter-dominated Universe), and the luminosity distance per unit co-moving radius increases $\propto (1+z)$.  In addition, as we have seen in examining the redshift-evolution of the distribution functions, the density of high luminosity sources increases with redshift up to $z \simeq 0.5$--1, but decreases rapidly at greater redshift.

The decline in total merger rate toward greater distance is not, however, the whole story.  
As Figure~\ref{fig:distfctn} shows, the number of systems with relatively large accretion rate (${\dot m} \sim 0.1$) decreases slowly with distance, as expected in a cosmological context, where black hole activity decreases in strength as galaxies become progressively poorer in gas.  On the other hand, the characteristic mass increases with distance; this may seem counterintuitive because black hole masses increase with time. The reason is related to the inclusion of the dynamical friction timescale. The lighter black holes that start their orbital decay at higher redshift have longer dynamical friction timescales, and therefore do not become bound in a binary until later than more massive black holes that start their orbital evolution at the same time. This combination of the evolution of binary mass and accretion rate implies that high luminosity sources ($L \propto \dot m M$) can be found out to rather greater distances than the average system.  

\begin{figure}
\begin{center}
\includegraphics[width=0.6\textwidth,angle=90]{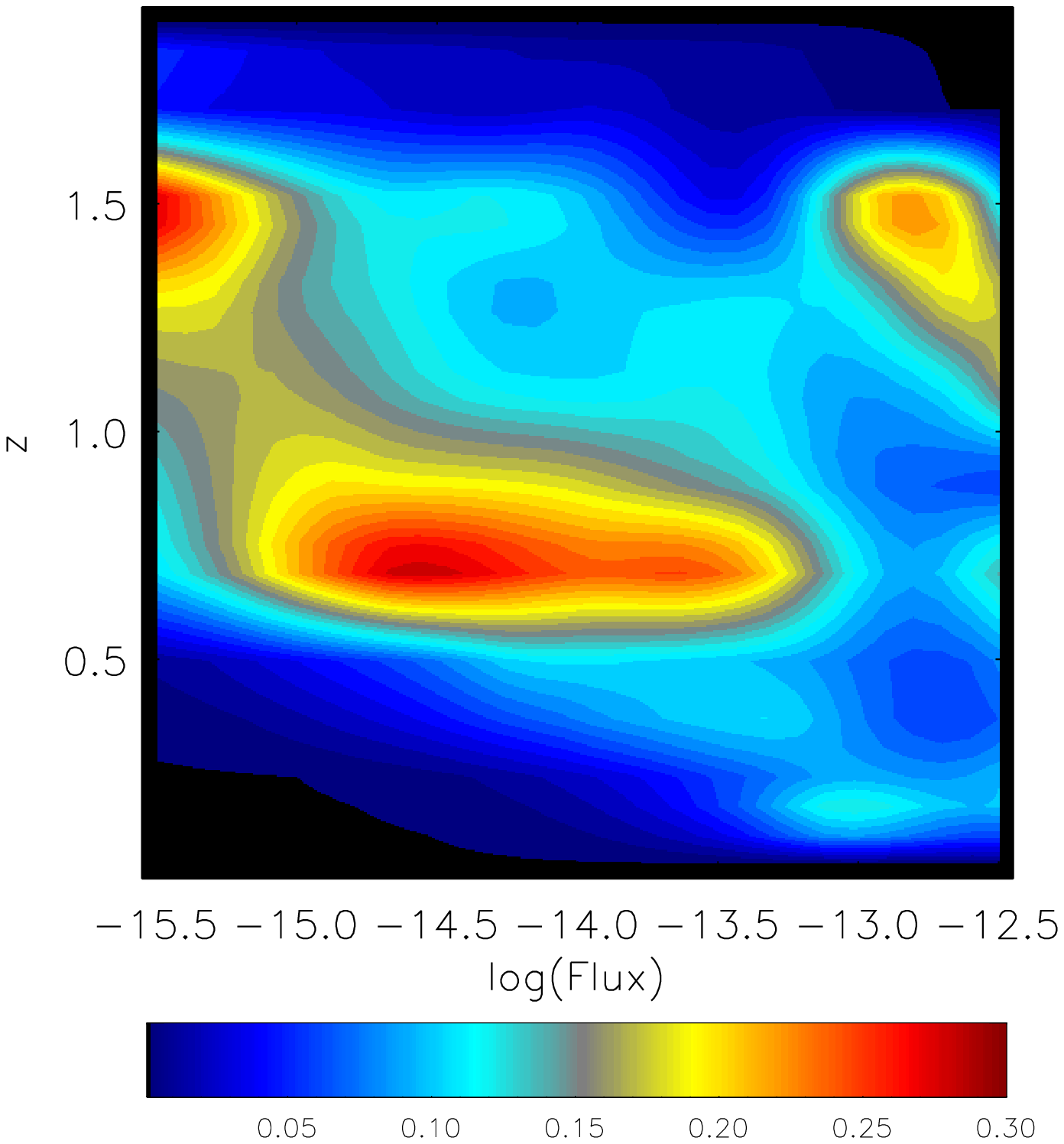}
\includegraphics[width=0.6\textwidth,angle=90]{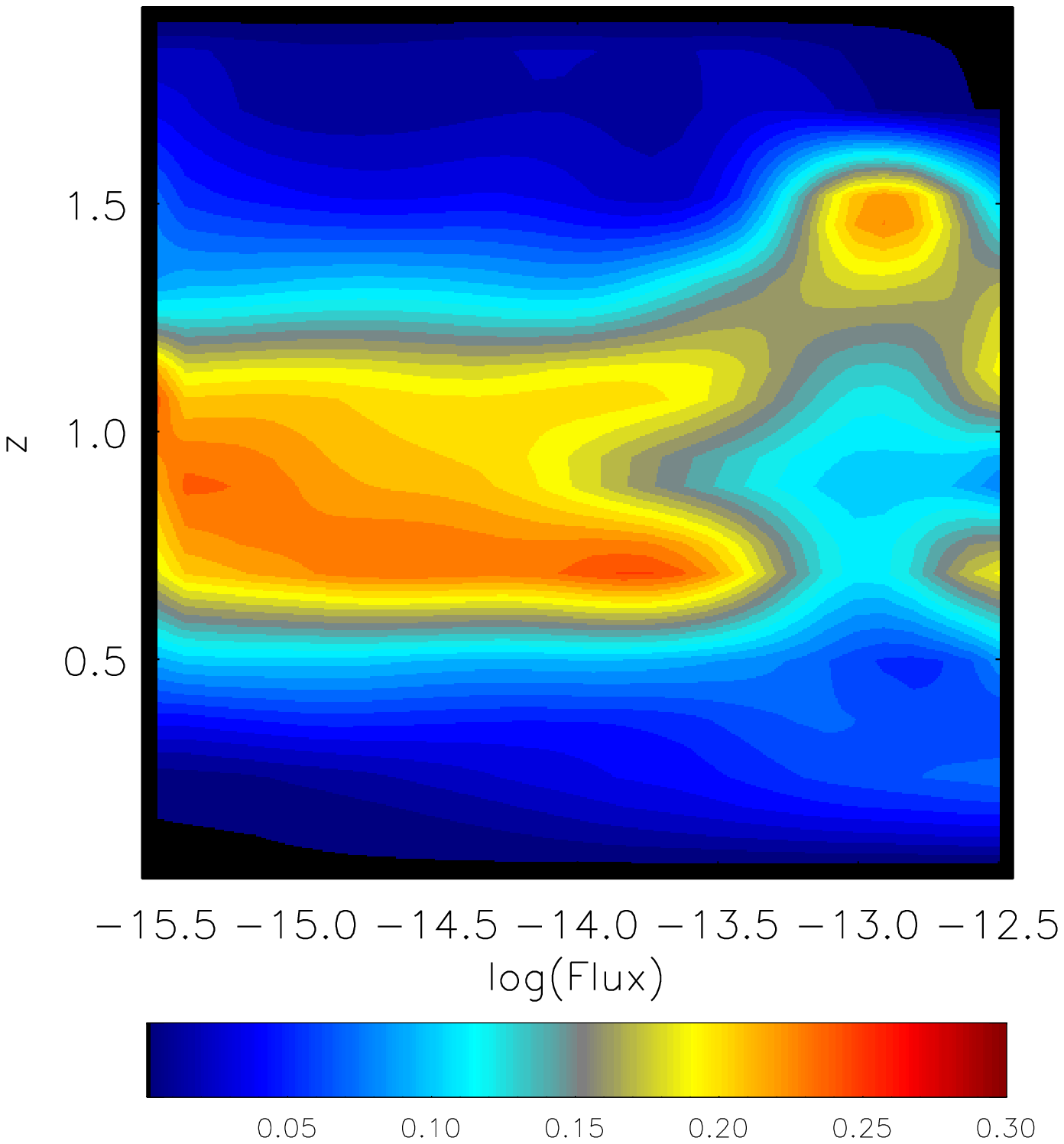}
\caption{Fractional distribution in redshift for the brighter end of the source count distribution with linear color contours.  (Top) X-ray signatures.  (Bottom) ``Notch" signatures.}
\label{fig:zdist}
\end{center}
\end{figure}

These trends are illustrated in a different way in Figure~\ref{fig:zdist}, where we plot $\frac{df}{dz}(F)$, the redshift probability density for a given flux.  For both signals, most of the potentially observable cases have redshifts in the range $0.6 \leq z \lesssim 1$, but the notch redshift distribution extends $\simeq 0.2$ higher in redshift than the X-ray distribution.  In both cases, the brightest flux objects are concentrated at $z \simeq 1.5$, but this may be an artifact of small-number statistics: these peaks can be attributed to the influence of only 2--3 binaries in the simulations.

\begin{figure}

\includegraphics[width=0.8\textwidth,angle=90]{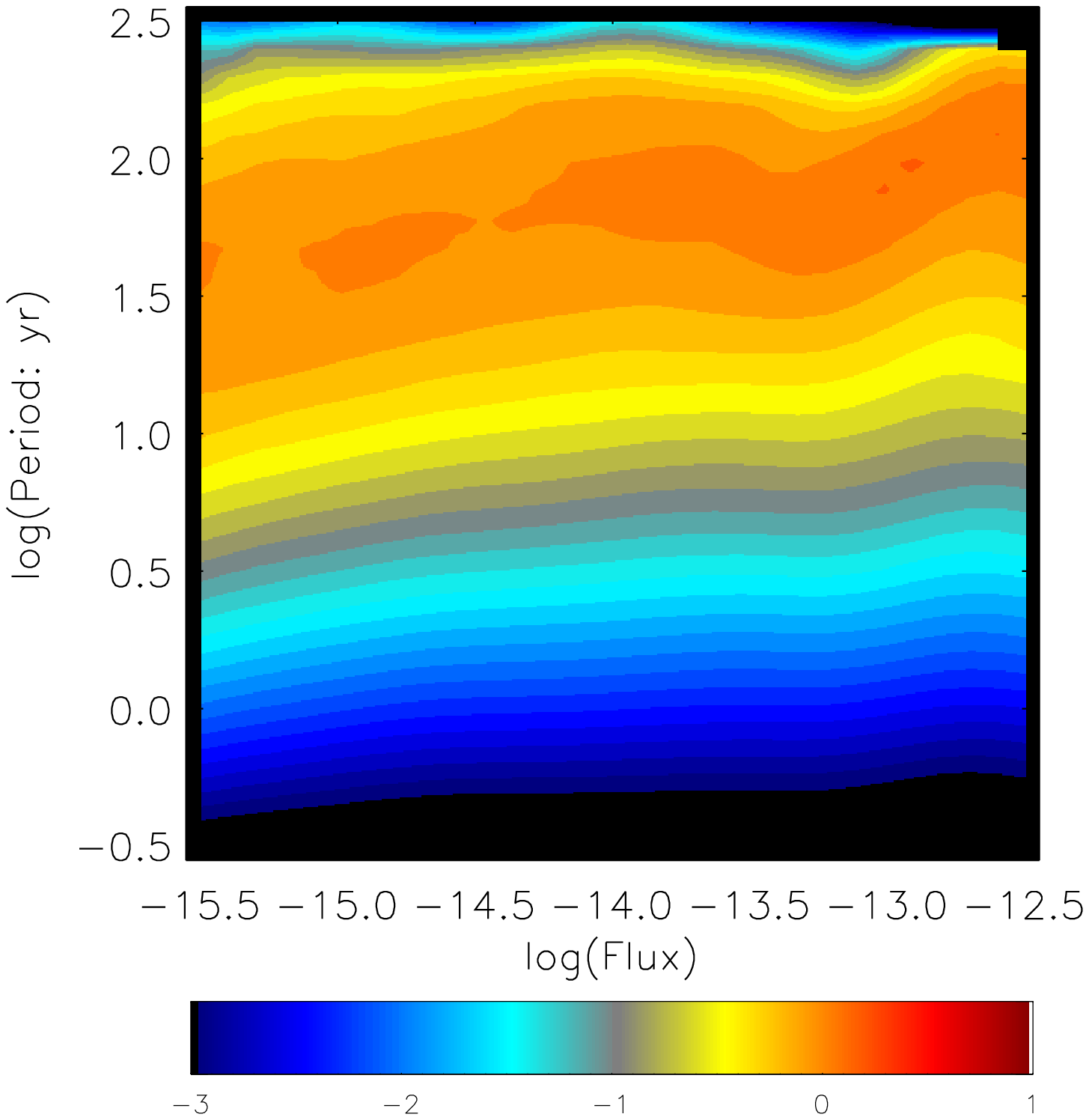}
\caption{Logarithm (base 10) of the orbital period probability density $df/d\ln P_{\rm orb}$ as a function of hard X-ray flux $F$.}
\label{fig:orbfrac}
\end{figure}

As already noted, our method also predicts the distribution of orbital periods found in association with these signatures.  In particular, the distribution of orbital periods (in the observed frame) as a function of flux can be described by the probability density of systems having a given period when they are detected with hard X-rays of flux $F$.  The quantity $[df/d\ln P_{\rm orb}](F)$ is displayed in Figure~\ref{fig:orbfrac}.  The range of orbital periods is capped at $\approx 300$~yr because we impose a maximum system mass of $3.16 \times 10^9M_\odot$ and a maximum separation $a/r_g = 10^3$.  Across the range of fluxes examined, the peak of the probability distribution is in the range $\simeq 50$--200~yr, rising slightly toward higher fluxes.   However, at all fluxes, there is a broad tail to shorter periods.   The fraction having periods more easily measured on human timescales (i.e., $< 5$~yr) is $\simeq 10\%$.

\section{Potential Observability}

Published large-area surveys in hard X-rays ($ > 2$~keV) (e.g., the RXTE [\citet{Revnivtsev2004}] and XMM [\citet{Warwick2012}] slew surveys) had flux limits $\sim 1 \times 10^{-11}$~erg~cm$^{-2}$~s$^{-1}$, well above the level at which we predict the numbers of SMBBHs.   However, the all-sky eRASS:8 survey planned for eROSITA (planned launch date mid-2019) is expected to have a flux limit at $1.6 \times 10^{-13}$~erg~cm$^{-2}$~s$^{-1}$ for point sources in the 2--10~keV band \citep{eRosita}.  Thus, eROSITA could potentially see large numbers of these objects if they could be distinguished from ordinary AGN either by possessing unusually hard spectra or by exhibiting periodic modulation.

There is a similarly positive prospect for discovering ``notch" sources in the near future.  A V-band flux $\simeq 10^{-13}$~erg~cm$^{-2}$~s$^{-1}$ corresponds to $m_V \simeq 18$, 1--2 magnitudes brighter than, for example, the original SDSS flux limit for quasar spectroscopy.  Thus, even within the original SDSS Quasar Survey (covering $\simeq 1/4$ of the sky), there might be $\sim 100$ ``notch" sources.  Moreover, the plan for SDSS-V \citep{SDSSV} includes coordination of its survey with that of eROSITA: an SDSS-V spectrum will be acquired for every accreting supermassive black hole identified by eROSITA.  Thus, it will be possible to examine all the objects in this shared sample for both signals. There is, however, the additional difficulty that by virtue of the notch's existence, the observable flux of the system is diminished; if the notch edge occurs in the middle of the spectrum and $\nu F_\nu$ is flat across the range of the spectrum, the observed magnitude is made $\simeq 0.7$~mag fainter.   

Compared to the general population at the same redshift and in the same flux range, binaries have higher masses. This is the result of two effects.  First, a higher merger rate for more massive halos, which host more massive galaxies and therefore more massive black holes, implies a larger primary black hole mass.  Second, the dynamical friction timescale is inversely proportional to the mass of the infalling black hole, so binary formation preferentially selects relatively massive secondary black holes.  By contrast, the accretion rate, compared at the same redshift and in the same flux range, is lower than in the general population, except at $z<0.6$.  This fact follows from binaries being typically very massive: at a given redshift the mean accretion rate decreases for increasing BH mass \citep[see Fig.~14 in][]{V2016}. If information on the host galaxies is available, our model suggests therefore that at a given flux level for an AGN, the most massive galaxies have a higher probability of hosting a binary, but they may be somewhat fainter than the mean isolated AGN in the same sort of galaxy.

The simulation adopted here is not ideal for studying LISA sources because the sensitivity curve peaks at masses $<10^7 \, M_\odot$, although it should be able to see events with mass in our range (e.g., $M_{\rm total} = 5\times 10^7 \, M_\odot$ with $q=0.2$ out to $z\sim 10$: \citealp{2017arXiv170200786A}).  Simulations with finer mass and spatial resolution are therefore required in order to define the population of LISA sources having these EM signatures.  For this reason, we have also conducted a smaller-volume, higher-resolution simulation whose black hole mass range extends to $\lesssim 10^4 M_\odot$ (Dubois et al. in prep.).   Because of its smaller volume (a sphere 10~Mpc in co-moving radius), it contains only $\sim 10$ supermassive black hole binaries per redshift slice, and therefore suffers badly from the fluctuations of small samples. It also extends only  to $z=0.6$, so it cannot provide any nearby systems.  Nonetheless, analysis of the sort we carried out on our primary, higher-mass, sample indicates that there could be $\gtrsim 100$ binaries with masses in the LISA range and notch or X-ray signals with flux $\sim 10^{-15}$~erg~cm$^{-3}$~s$^{-1}$.  An extension of this simulation to lower redshift would likely populate the higher-flux range at some level.

On the other hand, the binaries we have analyzed in this paper, with their high masses ($>10^7 \, M_\odot$) and low redshift ($z<2$), are ideal counterparts for PTA experiments.  The range of orbital periods for PTA targets overlaps significantly with the range for our systems: a period of 1~yr corresponds to a system with $a \simeq 500 M_8^{-2/3} r_g$, and, as we have shown, although only a minority of the brighter systems will have periods this short, it is not a tiny minority.  Because candidate SMBBHs discovered through either the hard X-ray or ``notch" signatures would have very well-defined locations in the sky, and perhaps in some cases well-defined orbital periods, the PTA detection threshold should be considerably lower than in a blind search: the statistical quality of parametric fits is greatly improved by diminishing the volume of acceptable parameter space.  PTA sensitivity could also be enhanced by identifying additional pulsars in the directions most needed for better S/N on specific candidate systems.

\section{Conclusions}

We have employed simulations in which galaxy and supermassive black hole evolution are coupled to predict the population of supermassive binary black hole systems potentially identifiable through their distinct electromagnetic signatures.  In particular, we have focused on two such signatures---a ``notch" in the IR/optical/UV spectrum, and a periodically-modulated hard X-ray component---because they are features unique to {\it binary} accreting supermassive black holes, as opposed to single accreting supermassive black holes, i.e, ordinary AGN.   They are, in addition, most readily visible when the binary separation is relatively close, but not quite in the immediately pre-merger state, i.e., $15 \lesssim a/r_g \lesssim 10^3$ (to see the ``notch" signal, there is a further restriction on $a/r_g$, but its effect partially duplicates the primary constraint.  These separations correspond to coalescence times from $0.12 M_8$~yr to
$2.4 \times 10^6M_8$~yr.

Although these estimates are subject to a number of uncertainties, they indicate that there may be enough systems exhibiting one or both of these signals to make searches worthwhile.   There may be $\sim 10^2$ of each with fluxes in the bands containing the signal $\gtrsim 10^{-13}$~erg~cm$^{-2}$~s$^{-1}$.  This flux is well above the minimum for inclusion in, for example, the SDSS quasar spectroscopic sample, and it roughly matches the flux limit for the all-sky eRASS:8 survey planned for eROSITA.   The source count curves for both signals are relatively shallow ($d\ln N/d\ln F \sim -1$) because, due to the long time required to produce such a binary, they may be predominantly found at modest redshift ($0.5 \lesssim z \lesssim 1$).

Their period distribution also makes them susceptible to observation.  Although most are about an order of magnitude too long for convenient human study ($\sim 100$~yr), an interesting fraction ($\sim 10\%$) have periods $\lesssim 5$~yr.  The shorter period systems are also in the range that can be probed by PTA experiments, providing a potentially important synergy.

Thus, there is a reasonable prospect that searches based on spectral features will be the first avenue to unveil the supermassive binary black hole population.

 \section*{Acknowledgements}

This work was partially supported by NSF grants AST-1516299 and AST-1715032 (JHK).  It was also partially supported by a grant from the Simons Foundation (559794, JHK).  MV acknowledges funding from the European Research Council under the European Community's Seventh Framework Programme (FP7/2007-2013 Grant Agreement no.\ 614199, project ``BLACK''). MV thanks Siyuan Chen and Alberto Sesana, and JHK thanks Sarah Burke-Spolaor for discussions on PTA limits.  JHK also thanks the organizers of his colloquia at Hebrew University and Ben-Gurion University, where questions raised by audience-members stimulated his work on this project.  We are grateful to Cole Miller for asking several insightful questions.

\end{document}